\newif\ifreview
\newif\ifarxiv

\reviewfalse

\arxivtrue

\ifreview
    \documentclass[sigconf,anonymous,review]{acmart}
\else
    \ifarxiv
        \documentclass[sigconf,nonacm]{acmart}
    \else
        \documentclass[sigconf]{acmart}
    \fi
\fi

\usepackage{enumitem}
\usepackage{multirow,multicol,makecell}
\usepackage{tikz}
\usepackage{pgfplots}
\pgfplotsset{compat=1.18}
\usetikzlibrary{calc}
\usepackage[ruled]{algorithm2e}
\usepackage{tabularx, hhline}
\definecolor{grey}{gray}{0.7}
\usepackage{amsthm}
\newtheorem{theorem}{Theorem}

\theoremstyle{definition}
\newtheorem{definition}[theorem]{Definition}
\theoremstyle{remark}

\AtBeginDocument{%
  }

\setcopyright{acmlicensed}
\copyrightyear{2026}
\acmYear{2026}
\acmDOI{XXXXXXX.XXXXXXX}
\acmConference[Conference acronym 'XX]{Make sure to enter the correct
  conference title from your rights confirmation email}{June 03--05,
  2018}{Woodstock, NY}
\acmISBN{978-1-4503-XXXX-X/2018/06}

\begin{document}

\title[Practical Auditing of Differentially Private Graph Release Mechanisms]{Practical Auditing of Differentially Private Graph Release Mechanisms: Methodology, Toolkit, and Case Study}

\author{Nicholas D'Silva}
\orcid{0009-0008-6195-7662}
\affiliation{%
  \institution{The University of New South Wales}
  \city{Sydney}
  \state{NSW}
  \country{Australia}}
\email{n.dsilva@unsw.edu.au}

\author{Surya Nepal}
\orcid{0000-0002-3289-6599}
\affiliation{%
  \institution{CSIRO's Data61 \and The University of New South Wales}
  \city{Sydney}
  \state{NSW}
  \country{Australia}}
\email{surya.nepal@data61.csiro.au}

\author{Salil S. Kanhere}
\orcid{0000-0002-1835-3475}
\affiliation{%
  \institution{The University of New South Wales}
  \city{Sydney}
  \state{NSW}
  \country{Australia}}
\email{salil.kanhere@unsw.edu.au}

\renewcommand{\shortauthors}{D'Silva et al.}

\begin{abstract}
Differential Privacy (DP) has become a prominent approach for protecting graph data releases, yet empirical auditing methodologies have largely been developed for tabular datasets and machine learning pipelines rather than graph-structured outputs. We present a methodology for empirically auditing DP graph release mechanisms, identifying two graph-specific audit design problems: canary construction and witness statistic selection. We combine these components with an established statistically certified testing procedure and implement the resulting methodology as an open-source Python toolkit. To demonstrate the approach, we audit a published state-of-the-art Node-DP graph release mechanism. Across a series of synthetic and real-world canary pairs, we obtain certified lower bounds on the privacy parameter $\varepsilon$ that exceed the claimed value, yielding counterexamples sufficient to falsify the claimed guarantee for the audited implementation. The violations persist under progressively stronger restrictions on the input domain, including non-empty and connected graphs. We also show that witness selection can materially affect whether a violation is detected, highlighting the importance of evaluating a diverse battery of statistics when auditing DP graph release mechanisms.
\end{abstract}




\maketitle

\section{Introduction} \label{introduction}
Graph data is valued for its ability to represent rich structural and relational information, with applications spanning domains such as social networks, infrastructure, and healthcare~\cite{nr}. While such data offers substantial analytical value, it also raises significant privacy concerns as the structure may encode sensitive relationships, entities, or identifying information~\cite{Narayanan09, Korolova08, Yuan24}. Differential Privacy~\cite{Dwork06} (DP) has become a leading approach for protecting graph data releases by providing formal, quantifiable privacy guarantees.

However, applying DP to graphs remains challenging in practice. Unlike tabular data, graph elements are structurally related, giving rise to multiple graph-specific privacy definitions, including Edge-DP~\cite{Karwa11}, which aims to protect edges, and Node-DP~\cite{Kasiviswanathan13}, which aims to protect nodes, as well as a diverse collection of DP graph release mechanisms. A large body of work therefore focuses on designing mechanisms that provide favourable privacy--utility trade-offs under these various graph privacy definitions~\cite{Li23}.

In recent years, empirical auditing has emerged as a complementary approach to evaluating DP mechanisms. Rather than attempting to prove a privacy guarantee mathematically, an empirical audit seeks statistical evidence that a mechanism violates its claimed DP guarantee. For a selected pair of neighbouring inputs, the auditor samples outputs and tests whether a statistical decision rule can distinguish the inputs in a way that is incompatible with the claimed $(\varepsilon,\delta)$-DP bound. Detecting such a violation therefore provides a statistically certified counterexample demonstrating that the implemented mechanism behaves inconsistently with its advertised guarantee~\cite{Jagielski20, Nasr21, Steinke23}. Such techniques have been successfully applied to machine learning systems and have revealed implementation flaws, privacy-accounting mistakes, and other discrepancies between theoretical guarantees and deployed behaviour~\cite{Tramer22}. However, existing auditing methodologies have largely been developed for tabular datasets and machine learning pipelines, where the privacy-relevant unit typically corresponds to a self-contained record or training example.

The graph release setting introduces additional challenges. Under graph DP definitions, the perturbation distinguishing neighbouring datasets is structurally embedded within a network of dependencies. Thus, auditors must address two distinct design problems. First, they must construct canaries whose structural context is meaningful for the audited graph mechanism. Second, they must identify witness statistics that can expose resulting distributional differences in released graphs. Unlike many prior tabular data or machine learning DP audits, DP graph releases do not admit a single natural attack signal, and different witness statistics may reveal fundamentally different aspects of graph structure. 

In this work, we propose a methodology for empirically auditing DP graph release mechanisms. It separates canary construction from witness statistic selection and combines these graph-specific components with established statistical certification techniques from the DP auditing literature~\cite{Jagielski20}. We implement the procedure as an open-source Python toolkit that supports user-defined graph releases, enabling adaptability across different neighbouring relations, canary construction procedures, and witness statistics.

To demonstrate the methodology, we audit the authors' released implementation of PrivDPR~\cite{Zhang25}, a recent state-of-the-art Node-DP graph release mechanism reported to achieve strong utility in both the original publication and subsequent independent evaluations~\cite{Dsilva26}. Our evaluation begins with a trivial edge-case counterexample and proceeds through a series of increasingly restrictive synthetic canaries before considering real-world graph datasets. Across all audited canaries, we obtain statistically certified privacy lower bounds exceeding the claimed privacy budget. The resulting violations persist under restrictions that favour the audited mechanism, including non-empty and connected input graphs.

\paragraph{Contributions} To summarise, our contributions are as follows:
\begin{enumerate}[leftmargin=*]
    \item We formulate a methodology for empirically auditing DP graph release mechanisms. We identify two graph-specific audit design problems -- canary construction and witness statistic selection -- that arise because a local graph change can affect structure throughout the release.
    \item We implement this methodology as an open-source Python toolkit\footnote{\label{fn:repo}\ifreview\url{https://github.com/ANONYMISED_FOR_REVIEW}\else\url{https://github.com/ndsi6382/DP-GRM_Auditor}\fi} supporting user-defined graph release mechanisms, neighbouring relations, canary construction procedures, and witness statistics.
    \item We audit the authors' released implementation of PrivDPR~\cite{Zhang25}, a recent state-of-the-art Node-DP graph release mechanism. Across a series of synthetic and real-world canary pairs, we obtain statistically certified lower bounds that exceed the claimed privacy budget, thereby falsifying the claimed guarantee for the audited implementation.
    \item We show that violation detection depends on witness selection for graphs, empirically motivating the use of a diverse battery of statistics when auditing DP graph release mechanisms. 
\end{enumerate}

\paragraph{Organisation} We present foundational and related work in Sec.~\ref{related-work}. We detail the audit methodology in Sec.~\ref{audit}, and discuss practical considerations in Sec.~\ref{practical-considerations}. The case study on PrivDPR is presented in Sec.~\ref{case-study-privdpr}. Limitations and future work are discussed in Sec.~\ref{future-work}, with conclusions provided in Sec.~\ref{conclusion}.

\section{Background and Related Work}\label{related-work}
\subsection{Background on Differential Privacy}
Differential Privacy~\cite{Dwork06} (DP) provides formal and quantifiable privacy guarantees for computations over sensitive datasets. Intuitively, outputs should be hard to distinguish between neighbouring datasets, limiting inferable information about individual records.

\begin{definition}[Differential Privacy~\cite{Dwork06}]
A randomised algorithm $A : \mathcal{X} \to \mathcal{Y}$ is $(\varepsilon,\delta)$-\textit{differentially private} if, for all neighbouring datasets $X,X' \in \mathcal{X}$ and all measurable events $Y \subseteq \mathcal{Y}$,
\begin{equation}\label{eq:differential-privacy}
    \Pr[A(X) \in Y]
    \le e^\varepsilon \Pr[A(X') \in Y] + \delta.
\end{equation} 
\end{definition}
\noindent $\varepsilon > 0$ is the privacy budget and $\delta \in [0,1)$ is the additive relaxation parameter. Smaller $\varepsilon$ provides stronger privacy guarantees, generally at the cost of utility. For a query $f$, its global sensitivity is given by $\max_{X,X'} \|f(X)-f(X')\|$, where $X, X'$ are neighbouring datasets. Sensitivity determines the magnitude of noise many DP mechanisms require and therefore directly affects the privacy--utility trade-off. When $\delta=0$, the guarantee is called \textit{$\varepsilon$-Pure DP}; otherwise, it is
\textit{$(\varepsilon,\delta)$-Approximate DP}.

DP expresses its privacy target via the neighbouring dataset relation. There are two typical forms of such relations: unbounded, where neighbouring datasets are obtained by adding or removing a record, protecting whether an individual is in the dataset; and bounded, where the relation is defined by substituting a record, thus providing attribute privacy under the condition that an individual is already assumed to be included in the dataset. Bounded DP is generally viewed as a weaker privacy guarantee than unbounded when the same $\varepsilon$ is used, since it excludes membership changes from the neighbouring relation~\cite{Dwork14}.

When applying DP to graphs, neighbouring relations are typically defined over either the edge set, providing Edge-DP~\cite{Karwa11}, or the vertex set, providing Node-DP~\cite{Kasiviswanathan13}. Edge-DP is typically instantiated under the unbounded notion, while Node-DP mechanisms have been developed over both bounded and unbounded variants.

\subsection{Related Work}
Auditing DP mechanisms is important for ensuring that real-world deployments provide their claimed privacy guarantees in practice, motivating research into effective auditing methods. Auditing exploits the gap between a mechanism's proven guarantee and its behaviour in practice. Ding et al.~\cite{Ding18} formalised this as statistical counterexample detection, while Bichsel et al.~\cite{Bichsel18, Bichsel21} develop methods to automate counterexample and classifier discovery. A prominent methodology for auditing DP-SGD was established by Jagielski et al.~\cite{Jagielski20} and was subsequently extended by Nasr et al.~\cite{Nasr21, Nasr23} and Steinke et al.~\cite{Steinke23}. We adopt this methodology directly, while our contribution concerns what precedes it, namely additional considerations for applying it to the graph domain. Within the release setting, Annamalai et al.~\cite{Annamalai24} audit DP synthetic data generators, the tabular analogue of our problem, and show that empirical leakage can exceed loose theoretical bounds. The work assumes canonical statistics transfer across mechanisms sharing an architecture, an assumption that the graph-specific mechanisms we consider here do not necessarily satisfy.

Our audits target DP graph release mechanisms. Existing surveys consolidate this literature~\cite{Li23}, as well as the broader pre-DP anonymisation space~\cite{Casas-Roma17}. Other works also benchmark state-of-the-art mechanisms~\cite{Liu25} and examine practical considerations for deployment~\cite {Dsilva26}. The present work complements this evaluation axis methodologically and can be used alongside such evaluation frameworks to enhance their comprehensiveness.

It is common to use graph de-anonymisation~\cite{Pedarsani13, Nilizadeh14}, reconstruction~\cite{Azogagh25}, and inference attacks~\cite{Korolova08, Yuan24} to quantify empirical privacy preservation on DP released graphs~\cite{Dsilva26}. Our case study shows that such attacks can miss a certified violation that our audit can detect directly. An attack asks whether an output is exploitable, whereas an audit asks whether the mechanism's output distribution matches its claimed DP guarantee. We view the two as complementary.

In contrast, no prior work adapts a principled empirical DP auditing methodology to the graph release setting, despite this being non-trivial due to additional structural and relational considerations. While we do not claim novelty in the underlying statistical construction~\cite{Jagielski20}, our contribution is adapting its surrounding methodology to a setting where it cannot be inherited unmodified.

\section{Audit Methodology}\label{audit}
This section presents a black-box procedure for testing whether an implemented graph release mechanism is consistent with a claimed DP guarantee. The procedure is a falsification test that provides statistically certified evidence that a claimed guarantee is violated; however, an unsuccessful audit does not establish that the mechanism satisfies the guarantee in general.

\subsection{Problem Setting and Audit Objective}
Let $\mathcal{G}$ denote the domain of graphs and let $A: \mathcal{G} \to \mathcal{Y}$ be a randomised mechanism with output space $\mathcal{Y}$. In our graph synthesis setting, $\mathcal{Y}=\mathcal{G}$, although this method also applies when $A$ releases statistics, model parameters, or other graph-derived objects. 

If $A$ is claimed to satisfy $(\varepsilon_0,\delta)$-DP with respect to a specified neighbouring relation $\sim$, then for every pair $G\sim G'$ and all measurable events $Y\subseteq\mathcal{Y}$,
\begin{equation} 
\Pr[A(G)\in Y] \leq e^{\varepsilon_0}\Pr[A(G')\in Y]+\delta. 
\label{eq:dp-forward} 
\end{equation}
Since $\sim$ is assumed to be symmetric, the same inequality holds when exchanging $G$ and $G'$. Also note that the neighbouring relation is part of the audited claim and must be fixed before an audit. Depending on the mechanism, this may represent edge~\cite{Karwa11}, node~\cite{Kasiviswanathan13}, or another form of graph adjacency. 

We assume black-box query access to $A$, i.e., the auditor may execute the mechanism on chosen input graphs and observe its released outputs but does not inspect or modify its internal computation, consistent with other empirical audits of DP mechanisms~\cite{Jagielski20, Nasr21}. The claim being tested concerns the implemented mechanism, including any implementation-level behaviour that affects its observable output distribution.

For a fixed neighbouring pair $G \sim G'$, the audit searches for an event $Y$ for which the output probabilities under $A(G)$ and $A(G')$ are incompatible with the claimed DP parameters. If $\alpha = \Pr[A(G)\in Y]$ and $\beta =\Pr[A(G')\in Y]$, then Eq.~\eqref{eq:dp-forward} requires $\alpha \leq e^{\varepsilon_0} \beta + \delta$. Therefore, any statistically certified lower bound on $\alpha$ and upper bound on $\beta$ can be used to derive a lower bound on the value of $\varepsilon$ required for this event. 

As previously mentioned, auditing DP on graph-structured data remains challenging as established methods do not directly address dependencies induced by the graph structure. This motivates us to adapt prior auditing methods to the graph setting. For graphs, the auditor must reason not only about which element differs between neighbouring inputs, but also about how that element is situated within the surrounding graph and how its influence can propagate through the mechanism. Accordingly, graph-specific auditing requires explicit design choices regarding both neighbouring-pair construction with specified canary contexts, and output-based witness statistics that capture different structural properties.

\subsection{Graph Canaries}
An audit can disprove a universal DP claim using a single neighbouring pair, but its ability to find such a pair depends on how the pair is constructed. We refer to the graph element or structure that differs between $G$ and $G'$ as the \textit{canary}. For example, an Edge-DP canary may be an edge whose inclusion closes a triangle, whereas a Node-DP canary may be a node with a specified degree and neighbourhood structure.

Unlike a tabular record that can often be varied while leaving the remaining records unchanged, a graph element derives much of its meaning from its relationships with the surrounding graph. Two valid canaries under the same neighbouring relation may therefore interact very differently with a mechanism. A removed node, for example, may be isolated, highly connected, or positioned within a dense subgraph. These structural differences can substantially affect both the mechanism's treatment of the canary and the auditor's ability to distinguish the resulting output distributions. Therefore, auditing graph DP mechanisms requires explicit reasoning about the structural context in which a canary is embedded, not just its presence or absence.

To make this design choice explicit, let $\phi(x; G)$ denote an auditor-defined description of the structural context of a canary $x$ in graph $G$. Depending on the audited mechanism, $\phi$ may include degree, local density, centrality, or participation in selected substructures. For example, degree describes the local extent of a node perturbation, while the proportion of incident edges relative to the graph's total edge count describes its relative magnitude. Here, $\phi$ is a design descriptor that guides canary construction or selection, rather than an additional input to the statistical test.

The auditor may then form a finite collection $\mathcal{C}=\{(G_i,G_i')\}_{i=1}^{k}$, where $G_i\sim G_i'$, representing the structural contexts examined by the audit. Each pair is held fixed during witness selection and audited separately using the procedure below.

\subsection{Graph Witness Statistics}
For each released output $Y \in \mathcal{Y}$, the auditor computes a collection of real-valued witness statistics $\mathcal{S} = \{s_1, \ldots, s_\ell\}$, where $s_j: \mathcal{Y} \to \mathbb{R}$ for all $1 \le j \le \ell$. A witness statistic converts a potentially complex released object into a scalar score that may distinguish outputs generated from neighbouring inputs. For a graph release, possible witnesses include degree-based, distance-based, spectral, centrality, and subgraph statistics. Whether a particular selection of statistics is appropriate depends on the mechanism's output type and the privacy claim being tested.

Importantly, canary construction and witness selection serve different purposes. Canary construction determines where the input graphs differ, whereas a witness statistic determines how the resulting outputs are distinguished. Mechanism-specific knowledge may inform both choices, but selecting them separately makes the search procedure clearer and allows every candidate pair to be evaluated using the full collection of statistics. 

A second challenge arises because graph releases are inherently multi-faceted objects. Unlike model auditing, where quantities such as loss, confidence, or gradient-derived scores often provide natural attack signals, a released graph has no single canonical notion of discrepancy. Structural differences may be expressed through degree distributions, connectivity patterns, distance structure, centrality measures, spectral properties, or subgraph counts, among others. Consequently, a witness that is highly informative for one graph mechanism may miss a privacy violation exposed through a different structural property.

The collection of statistics should therefore contain witnesses that capture different properties of the released object rather than several minor variations of the same statistic. Nevertheless, the battery remains a finite search space. Failure to find a distinguishing witness means only that the tested canaries and statistics did not reveal a violation; it does not imply that no other neighbouring pair, statistic, or decision rule could. 

\subsection{Selection of a Distinguishing Test}
Having defined graph canaries and witness statistics, the remainder of the audit follows the statistical framework of Jagielski et al.~\cite{Jagielski20}, which we use to select and certify distinguishing tests. For a fixed candidate pair $(G_i, G_i')$, we independently execute $A$ on both inputs to obtain labelled output samples:
\begin{equation*}
    Y_{i,1},\ldots,Y_{i,n_{\mathrm{sel}}} \overset{\mathrm{i.i.d.}}{\sim} A(G_i) \quad \text{and}\quad Y'_{i,1},\ldots,Y'_{i,n'_{\mathrm{sel}}} \overset{\mathrm{i.i.d.}}{\sim} A(G_i').
\end{equation*}
These samples of outputs form one pair-level selection set. We use the same outputs to compute every witness statistic $s\in\mathcal S$. For each fixed $s$, we conduct a separate test, choosing its threshold $\tau \in \mathcal{T}_{i,s}$, score-tail direction $d \in \{+1, -1\}$, and privacy direction $q$ using this selection set. Define the event:
\begin{equation*}
    E_{s,\tau,d} = \{Y\in\mathcal{Y}:d\,s(Y)\geq d\,\tau\}. 
\end{equation*}
The corresponding binary classifier is $F_{s,\tau,d}(Y) = \mathbf{1}[Y\in E_{s,\tau,d}]$, which outputs 1 when the event $E_{s,\tau,d}$ occurs and 0 otherwise.

For a fixed pair $i$ and witness statistic $s$, candidate thresholds are generated directly from the statistic values observed in the selection set. Specifically, let $\mathcal{T}_{i,s} = \{+\infty\} \cup \{s(Y_{i,j})\}_{j} \cup \{s(Y'_{i,j})\}_{j} \cup \{-\infty\}$, where thresholds are evaluated in descending order. For $d=+1$, the thresholds $+\infty$ and $-\infty$ give the all-negative and all-positive classifiers, respectively; for $d=-1$, these roles are reversed.

For each fixed pair $(G_i,G_i')$ and statistic $s$, let $q\in\{\mathrm{fwd}, \mathrm{rev}\}$ denote the direction of the DP inequality being tested. We treat outputs from $A(G_i)$ as the positive class and outputs from $A(G_i')$ as the negative class. For each candidate event $E_{s,\tau,d}$, let $\widehat{\mathrm{TPR}}$ and $\widehat{\mathrm{FPR}}$ be its empirical true-positive and false-positive rates on the selection set. We compute:
\begin{equation*}
    \widehat{\varepsilon}_{\mathrm{fwd}} = \log\left(\frac{\widehat{\mathrm{TPR}}-\delta}{\widehat{\mathrm{FPR}}}\right),
    \qquad
    \widehat{\varepsilon}_{\mathrm{rev}} = \log\left(\frac{1-\widehat{\mathrm{FPR}}-\delta}{1-\widehat{\mathrm{TPR}}}
    \right).
\end{equation*}
Here, $d$ determines the score-tail defining the event, while $q$ determines which direction of the DP inequality is tested.

We discard candidate directions for which the corresponding estimate is undefined, according to the zero-denominator convention used by the implementation. The selected values are:
\begin{equation*}
    (\tau^*_{i,s},d^*_{i,s},q^*_{i,s}) = \arg\max_{\substack{\tau\in\mathcal{T}_{i,s},\,
                         d\in\{+1,-1\},\,
                         q\in\{\mathrm{fwd},\mathrm{rev}\}}}
    \widehat{\varepsilon}_{q}.
\end{equation*}
If candidates tie, we retain the first according to a prespecified search order over thresholds, score-tail directions, and privacy directions. The selected event is:
\begin{equation*}
    E^*_{i,s} := E_{s,\tau^*_{i,s},d^*_{i,s}}.
\end{equation*}

Note that this selection is exploratory. The threshold, score-tail direction, and privacy direction are chosen using the selection data. Those data therefore cannot also be used to construct the final confidence bound, as this would invalidate the stated coverage due to adaptive selection across tests~\cite{Cox75}. We repeat the procedure separately for each statistic and each canary pair; we do not maximise selection scores across statistics or pool evidence across tests.

\subsection{Evaluation}
After selection, the distinguishing event is fixed as $E^*_{i,s}$. The auditor then obtains a new evaluation set of independent samples:
\begin{equation*}
    Z_1,\ldots,Z_{n_\text{eval}} \overset{\mathrm{i.i.d.}}{\sim} A(G_i) \quad \text{and} \quad Z'_1,\ldots,Z'_{n'_\text{eval}} \overset{\mathrm{i.i.d.}}{\sim} A(G_i').
\end{equation*} 
These outputs form one pair-level evaluation set, independent of the selection set, and are reused to compute every witness statistic. They must not be used to construct candidate thresholds or select the statistic-specific threshold, score-tail direction, or privacy direction. Define:
\begin{equation*}
(\alpha,\beta)=
\begin{cases}
\bigl(\Pr[A(G_i)\in E^*_{i,s}],\,\Pr[A(G_i')\in E^*_{i,s}]\bigr),
& \text{if } q^*_{i,s}=\mathrm{fwd};\\
\bigl(\Pr[A(G_i')\notin E^*_{i,s}],\,\Pr[A(G_i)\notin E^*_{i,s}]\bigr),
& \text{if } q^*_{i,s}=\mathrm{rev}.
\end{cases}
\end{equation*}

We compute a one-sided Clopper--Pearson lower confidence bound $\alpha_\text{lower}$ for $\alpha$, and a one-sided Clopper--Pearson upper confidence bound $\beta_\text{upper}$ for $\beta$~\cite{Clopper34}. To obtain joint coverage of at least $1-\gamma$, each bound is constructed with individual confidence level $1-\gamma/2$. By the union bound, $\Pr[\alpha \ge \alpha_\text{lower} \ \land \ \beta \le \beta_\text{upper}] \ge 1-\gamma$. This confidence level applies individually to each certified pair--witness test.

If $\alpha_\text{lower} > \delta$, Eq.~\eqref{eq:dp-forward} implies the following empirical lower bound on the privacy budget $\varepsilon$:
\begin{equation}\label{eq:empirical-epsilon-lb}
    \widehat{\varepsilon}_\text{LB} = \max \left\{ 0, \log \left( \frac{\alpha_\text{lower} - \delta}{\beta_\text{upper}} \right) \right\}.
\end{equation}
If $\alpha_\text{lower} \le \delta$, then the selected event does not yield a positive lower bound through this inequality, and we set $\widehat{\varepsilon}_\text{LB} = 0$.

Eq.~\eqref{eq:empirical-epsilon-lb} follows directly from the hypothesis-testing interpretation of DP by Kairouz et al.~\cite{Kairouz15}. It is also justified by DP post-processing, as the binary decision $F_{s,\tau^*_{i,s},d^*_{i,s}}$ is a deterministic function of $A$'s output and therefore satisfies the same DP guarantee as $A$~\cite{Dwork14}.

We reject the claimed $(\varepsilon_0,\delta)$ guarantee when $\widehat{\varepsilon}_\text{LB} > \varepsilon_0$. With probability at least $1-\gamma$ over the test, this result shows that the implemented mechanism violates the claimed DP inequality for the audited neighbouring pair (canary), witness statistic, and the selected event. Since DP requires the inequality to hold for every neighbouring pair and every measurable event, one such certified counterexample is sufficient to falsify the universal claim.

Conversely, if the lower bound does not exceed $\varepsilon_0$, the result is inconclusive. It may reflect a valid implementation, but it may also result from an insufficiently informative canary, statistic, threshold, or evaluation sample size. The audit therefore provides evidence of violation when successful, but never a certificate of privacy when unsuccessful~\cite{Ding18}.

\section{Practical Considerations}\label{practical-considerations}
\subsection{Time Complexity}
As the proposed methodology is intended for practical auditing, we briefly analyse its computational complexity as a function of the audited mechanism, witness statistics, and audit sample sizes.

For a fixed canary pair $(G_i, G'_i)$, we assume without loss of generality that $G_i$ is the larger graph. Let $N_\text{sel}:= n_\text{sel} + n'_\text{sel}$ denote the number of selection samples generated from the neighbouring graphs, and $N_\text{eval}:= n_\text{eval} + n'_\text{eval}$ denote the number of evaluation samples generated from the neighbouring graphs. Let $T_A(G_i)$ denote an upper bound on the cost of a single execution of the audited mechanism $A$ on the canary pair, and let $T_s(G_i)$ denote the worst-case cost of evaluating witness statistic $s$.

The audit then yields a mechanism cost of $O((N_\text{sel} + N_\text{eval})T_A(G_i))$ across the trials. Witness statistics must also be evaluated on both the selection and evaluation outputs, yielding a cost of $O((N_\text{sel}+N_\text{eval})T_s(G_i))$ per statistic. The threshold set $\mathcal{T}_{i,s}$ contains $O(N_\text{sel})$ candidate thresholds, and for each threshold and direction, empirical true-positive and false-positive rates are computed over the selection set, requiring $O(N_\text{sel})$ work, resulting in $O(N_\text{sel}^2)$ time per witness statistic. The remaining operations are constant time, so the overall time complexity for auditing a single canary pair is $O((N_\text{sel} + N_\text{eval})T_A(G_i) + \sum_{s \in \mathcal{S}}((N_\text{sel} + N_\text{eval})T_s(G_i) + N_\text{sel}^2))$.

Thus, the audit scales linearly with the number of mechanism executions and witness statistics, while the threshold search is quadratic in the number of selection samples. In practice, however, this asymptotic term is typically dominated by the cost of executing the audited mechanism and evaluating graph-based witness statistics, both of which operate on full graph outputs. Consequently, the threshold search generally constitutes a comparatively small portion of the overall audit runtime.

\subsection{Integration into Broader Evaluation Frameworks}
The proposed audit complements broader frameworks for selecting, interpreting, and evaluating DP graph release mechanisms~\cite{Dsilva26, Liu25}. Such frameworks typically consider the claimed privacy guarantee alongside utility and computational requirements. However, interpreting the resulting privacy--utility trade-off presupposes that the implemented mechanism satisfies its claimed privacy guarantee.

We therefore propose empirical graph DP auditing as an additional validation stage between DP mechanism configuration and interpreting its privacy--utility performance. After selecting a candidate mechanism and configuration, the practitioner evaluates its utility using application-relevant graph statistics and empirically audits its claimed privacy guarantee. The resulting privacy, utility, and auditing evidence should then be interpreted jointly. In particular, utility achieved under a nominal privacy parameter should not be treated as representing that privacy--utility operating point when an audit certifies that the implemented mechanism is incompatible with the claimed guarantee.

At the same time, empirical auditing does not replace formal privacy analysis. Since an audit tests only a finite collection of neighbouring pairs, witness statistics, and decision rules, an unsuccessful audit cannot establish that the implementation satisfies DP in general. Auditing should instead be viewed as a falsification layer that complements formal analysis~\cite{Jagielski20, Nasr21, Ding18}. Its practical value is that it evaluates the observable behaviour of the implemented mechanism and can therefore reveal implementation defects, configuration errors, privacy-accounting mistakes, or other discrepancies between a formal claim and its implementation~\cite{Tramer22, Chourasia26}.

\subsection{Implementation Details}
We implement the DP graph release auditing methodology as an open-source Python module.\textsuperscript{\ref{fn:repo}} It uses NumPy~\cite{numpy} and SciPy~\cite{scipy} for numerical and statistical computation, Matplotlib for visualisation, Typeguard for runtime type checking, and tqdm for progress reporting. Its modular design separates mechanism execution, witness computation, test selection, statistical certification, and result export. This separation allows users to supply mechanism outputs and witness functions with minimal mechanism-specific integration code. The repository also provides API documentation and facilities for exporting and visualising audit results. 

We include executable examples demonstrating the complete audit workflow, using canary pairs constructed from Zachary's Karate Club~\cite{ZacharysKarateClub} as the dataset and the method used in our case study (Sec.~\ref{case-study-privdpr}), PrivDPR, as the mechanism. The examples show how to construct canaries, load mechanism outputs, define candidate witness statistics, and proceed with distinguishing test selections and their independent evaluations. We also release the preprocessing and neighbouring-pair construction code for the datasets used in the case study. This supports reproducibility as the instantiated neighbouring relation and the structural context of each canary form part of the audited privacy claim.

For graph-valued outputs, we provide an optional collection of witness statistics implemented using NetworkX~\cite{networkx}, including degree-based, distance-based, centrality, spectral, and subgraph statistics. The core auditing procedure does not require NetworkX. Users may instead provide any callable that maps a mechanism output to a real-valued score, allowing the toolkit to interoperate with alternative graph libraries and output representations. The supplied NetworkX witnesses therefore provide a reusable starting point; they are not exhaustive or universally appropriate.

\section{Case Study: PrivDPR}\label{case-study-privdpr}
To demonstrate our audit, we instantiate it on PrivDPR as a case study. PrivDPR~\cite{Zhang25} is a Node-DP graph release mechanism that trains a deep neural network to approximate per-node PageRank scores, using an embedding matrix optimised with DP-noised gradients. A synthetic graph is then reconstructed by sampling edges proportional to scores derived from the learned embeddings.

The neighbouring relation used by PrivDPR is that of bounded node adjacency~\cite{Kearns16}. Here, two graphs are considered neighbouring if one graph can be obtained from the other by arbitrarily rewiring the edges incident to a single vertex. Formally, for $G = (V, E), G' = (V, E') \in \mathcal{G}_n$, $G \sim G'$ if and only if $\exists v \in V : E \setminus E_v = E' \setminus E'_v$, where $E_v = \{(v, u) \in E\} \cup \{(u, v) \in E\}$ is the set of all edges incident to $v$ in $G$ (and $E_v'$ analogously in $G'$). Note that the cardinality of $V$ and the identity of its elements are fixed across $ G$ and $G'$.

Empirical results presented in the original work~\cite{Zhang25}, as well as independent evaluations~\cite{Dsilva26}, indicate that PrivDPR is able to retain high levels of utility, far exceeding that of other existing Node-DP mechanisms, even comparable to that of some Edge-DP mechanisms. However, the challenge of retaining utility while satisfying Node-DP is widely acknowledged in the graph privacy literature~\cite{Kasiviswanathan13, Blocki13, Raskhodnikova16}. This motivates us to investigate the DP guarantees that the mechanism provides in practice.

We use the authors' publicly released implementation of the mechanism. The audit follows the reproducible black-box methodology described in Sec.~\ref{audit} and relies solely on publicly available artefacts, documentation, and configuration details, thereby supporting independent verification of the reported results. For compatibility, the implementation was migrated to TensorFlow~2~\cite{tensorflow} using TensorFlow's official \texttt{tf\_upgrade\_v2} conversion tool. This affects only TensorFlow API calls and does not alter the underlying PrivDPR algorithm. The same parameters were used as in the original work, including $\varepsilon = 0.1$ and $\delta = 10^{-5}$. For each execution, we run 200 selection trials and 200 evaluation trials, with $\gamma = 0.01$.

We evaluated a battery of witness statistics for each audited canary pair. For brevity, we present results for edge count, a representative witness statistic that consistently exposed the same qualitative behaviour observed across the broader statistic collection. This does not affect the validity of the falsification test -- because DP must hold for every measurable event, a statistically certified violation obtained using any single witness statistic is sufficient to falsify the claimed guarantee. The purpose of the witness battery is to improve the auditor's ability to discover a counterexample, rather than to aggregate evidence across statistics. The reported confidence bounds are pointwise; conditional on the selection data, each fixed canary-pair--witness-statistic test has coverage of at least $1-\gamma$ over its independent evaluation samples. We do not claim simultaneous coverage across all tested pairs and statistics. For completeness, we additionally show two representative non-falsifying statistics, to emphasise that witness selection can materially influence whether a violation is detected.

\subsection{Datasets and Canaries}
We begin with a trivial counterexample. Let $G$ be a $k$-star, and let $G'$ be the neighbouring graph obtained by isolating the central vertex, leaving $G'$ entirely disconnected. In our experiments, PrivDPR consistently produces an edgeless output on $G'$, while executions on $G$ yield non-empty outputs. The neighbouring inputs are therefore directly distinguishable from the released graph. Although this observation is sufficient to invalidate the universal DP claim, it relies on a degenerate graph construction and offers limited insight into the mechanism's behaviour on more representative inputs. One could, for example, reasonably argue that this edge-case need not be handled, as graph datasets are typically not empty.

We therefore focus our empirical evaluation on non-degenerate canary pairs that more informatively assess the robustness of the observed violation. Specifically, we construct four canary pairs that satisfy progressively stronger restrictions on the input domain. Although these restrictions are not required by the original privacy claim, we introduce them as increasingly favourable conditions for the audited mechanism. Examples include restricting attention to non-empty and connected graphs. Conceptually, the canaries may be viewed as belonging to a nested sequence of graph classes $\mathcal{G}_0 \supseteq \mathcal{G}_1 \supseteq \mathcal{G}_2 \supseteq \mathcal{G}_3$, where each successive class imposes additional constraints on the admissible inputs. After evaluating these progressively restricted synthetic constructions, we additionally derive canary pairs from two real-world datasets to assess whether analogous violations are observable outside controlled examples. The details of all neighbouring graph pairs that were empirically audited are provided in Table~\ref{tab:dataset}.

\subsubsection{Synthetic Constructions}
\paragraph{Construction A} The first restriction we impose is that the graph is non-empty. Let $G$ be a star with a central vertex $v$, $999$ leaves, and an additional edge $(u, w)$ between two of the leaf vertices $u, w$. Let $G'$ be obtained from $G$ by isolating $v$. Note that both $G$ and $G'$ are non-empty. A diagram of this construction is shown in Fig.~\ref{fig:construction-a}. 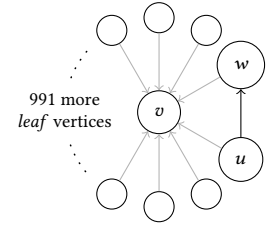
\begin{figure}
    \centering
    \begin{tikzpicture}[font=\footnotesize]
        \node[circle, draw, inner sep=4pt] (c) at (0,0) {$v$};
        \foreach \i/\angle/\lbl in {1/30/$w$,2/330/$u$} {
            \node[circle,draw,inner sep=4pt] (v\i)
                at (\angle:1.25cm) {\lbl};
            \draw[<-,grey] (c) -- (v\i);
        }
        \draw[<-] (v1) -- (v2);
        \foreach \angle in {60,90,120,300,270,240} {
            \node[circle,draw,inner sep=4pt] (v\angle)
                at (\angle:1.25cm) {};
            \draw[<-, grey] (c) -- (v\angle);
        }
        \draw[line cap=round, dash pattern=on 0.25pt off 3.5pt, line width=0.6pt] (140:1.25cm) arc (140:220:1.25cm);
        \node[rectangle, fill=white, align=center] 
            at (180:1.25cm) {$991$ more\\\textit{leaf} vertices};
    \end{tikzpicture}
    \caption{Construction A. Edges incident to $\boldsymbol{v}$ are rewired such that those in grey are not present in the neighbouring graph.}
    \label{fig:construction-a}
\end{figure}

\begin{figure}
    \centering
    \begin{tikzpicture}[font=\footnotesize]
        \node[circle, draw, inner sep=4pt] (c) at (0,0) {$v$};
        \foreach \i/\angle/\lbl in {1/30/$w$,2/330/$u$} {
            \node[circle,draw,inner sep=4pt] (v\i)
                at (\angle:1.25cm) {\lbl};
            \draw[<-, grey] (c) -- (v\i);
        }
        \draw[<-] (v1) -- (v2);
    
        \node[circle] (v1ext) at ($(v1)+(45:1.25cm)$) {};
        \draw[dashed] (v1) -- (v1ext);
        \node[circle] (v2ext) at ($(v2)+(315:1.25cm)$) {};
        \draw[dashed] (v2) -- (v2ext);
        \node[rectangle, fill=white, align=center] 
            at (0:2.5cm) {998 more vertices\\of an Erd\H{o}s--R\'enyi\\random graph};
        
        \foreach \angle in {60,90,120,300,270,240} {
            \node[circle,draw,inner sep=4pt] (v\angle)
                at (\angle:1.25cm) {};
            \draw[<-, grey] (c) -- (v\angle);
        }
        \draw[line cap=round, dash pattern=on 0.25pt off 3.5pt, line width=0.6pt] (140:1.25cm) arc (140:220:1.25cm);
        \node[rectangle, fill=white, align=center] 
            at (180:1.25cm) {$991$ more\\\textit{leaf} vertices};
    \end{tikzpicture}
    \caption{Construction B. Edges incident to $\boldsymbol{v}$ are rewired such that those in grey are not present in the neighbouring graph.}
    \label{fig:construction-b}
\end{figure}
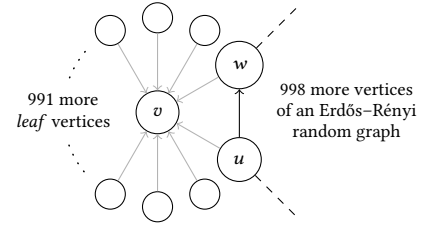

\paragraph{Construction B} We next investigate whether the violation persists when the structural perturbation constitutes only a small fraction of the overall graph. To do this, we modify Construction A so that the difference in edge count between $G$ and $G'$ comprises less than 10\% of the edges in $G$. Let $G$ be as per Construction A, except that the edge $(u, w)$ forms the intersection between the modified $k$-star and an Erd\H{o}s--R\'enyi random graph~\cite{Erdos59} on $n=1000$ vertices and $p=0.025$ edge probability. We obtain $G'$ by isolating $v$. A diagram of this construction is shown in Fig.~\ref{fig:construction-b}. 

\paragraph{Construction C} We next restrict our attention to connected graphs. Let $G$ be a star with a central vertex $v$, $999$ leaves, and an additional vertex $w$, connected to $v$ and all its leaf vertices. We obtain $G'$ by isolating $v$ from its leaves, but not from $w$. Note that both $G$ and $G'$ are connected. A diagram of this construction is shown in Fig.~\ref{fig:construction-c}. 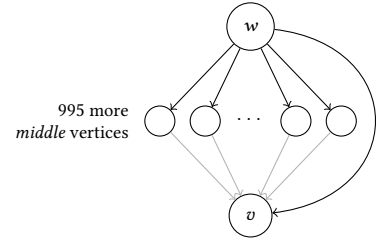
\begin{figure}
    \centering
    \begin{tikzpicture}[vertex/.style={circle, draw, inner sep=4pt}, font=\footnotesize]
        \node[vertex] (top) at (0, 1.25) {$w$};
        \node[vertex] (bottom) at (0,-1.25) {$v$};
        \foreach \i in {-2,-1,1,2} {
            \node[vertex] (mid\i) at (\i*0.6,0) {};
            \draw[->] (top) -- (mid\i);
            \draw[->, grey] (mid\i) -- (bottom);
        }
        \node at (0,0) {$\cdots$};
        \node[left, align=right] at (-1.5,0) {$995$ more\\\textit{middle} vertices};
        \draw[->] (top) .. controls (2.1,1) and (2.1,-1) .. (bottom);
    \end{tikzpicture}
    \caption{Construction C. Edges incident to $\boldsymbol{v}$ are rewired such that those in grey are not present in the neighbouring graph.}
    \label{fig:construction-c}
\end{figure}

\begin{figure}
    \centering
    \begin{tikzpicture}[vertex/.style={circle, draw, inner sep=4pt}, font=\footnotesize]
        \node[vertex] (top) at (0,1.25) {$w$};
        \node[circle] (topext) at ($(top)+(4:2.3cm)$) {};
        \draw[dashed] (top) -- (topext);
        \node[vertex] (bottom) at (0,-1.25) {$v$};
        \node[circle] (bottomext) at ($(bottom)+(356:2.3cm)$) {};
        \draw[dashed] (bottom) -- (bottomext);
        \node[rectangle, fill=white, align=center] 
            at (3,0) {998 more vertices\\of an Erd\H{o}s--R\'enyi\\random graph};
        \foreach \i in {-2,-1,1,2} {
            \node[vertex] (mid\i) at (\i*0.6,0) {};
            \draw[->] (top) -- (mid\i);
            \draw[->, grey] (mid\i) -- (bottom);
        }
        \node at (0,0) {$\cdots$};
        \node[left, align=right] at (-1.5,0) {$995$ more\\\textit{middle} vertices};
        \draw[->] (top) .. controls (2,1) and (2,-1) .. (bottom);
    \end{tikzpicture}
    \caption{Construction D. Edges incident to $\boldsymbol{v}$ are rewired such that those in grey are not present in the neighbouring graph.}
    \label{fig:construction-d}
\end{figure}
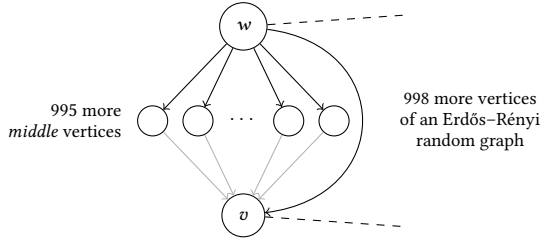

\paragraph{Construction D} Finally, we combine the restrictions of Constructions B and C by considering a connected graph in which the perturbation forms only a small fraction of the overall structure. Specifically, $G$ and $G'$ are as per Construction C, except that the edge $(w,v)$ forms the intersection between the current structure and another Erd\H{o}s--R\'enyi random graph on $n=1000$ vertices and $p=0.025$ edge probability. A diagram of this construction is shown in Fig.~\ref{fig:construction-d}. 

Collectively, Constructions A--D allow us to assess whether the observed
violation persists under increasingly restrictive and favourable
conditions for the audited mechanism.

\subsubsection{Real-World Datasets}
Constructing canary pairs from real-world datasets emphasises the practical implications of DP guarantees. While the previous constructions were controlled synthetic examples, we additionally demonstrate that analogous violations can be detected on real-world graph data. For example, similar star-shaped (or hub-and-spoke) structures to our counterexample arise naturally in many social network settings~\cite{Charbey19}. We therefore source two real-world datasets.

\paragraph{RT-LOLGOP~\cite{Lolgop}.} This dataset represents a Retweet network. We take the original dataset as $G$, and obtain $G'$ by isolating the maximum-degree vertex.

\paragraph{Anybeat~\cite{Anybeat}.} This dataset represents interactions taking place on the music-based online social network of the same name. Similarly, we take the original dataset as $G$, and obtain $G'$ by isolating the maximum-degree vertex.

\begin{table*}
    \centering
    \setlength{\tabcolsep}{3pt}
    \begin{tabularx}{\textwidth}{|r||*{12}{>{\centering\arraybackslash}X|}}
        \hline & \multicolumn{2}{c|}{\textbf{Construction A}} & \multicolumn{2}{c|}{\textbf{Construction B}} & \multicolumn{2}{c|}{\textbf{Construction C}} & \multicolumn{2}{c|}{\textbf{Construction D}} & \multicolumn{2}{c|}{\textbf{RT-LOLGOP}~\cite{Lolgop}} & \multicolumn{2}{c|}{\textbf{Anybeat}~\cite{Anybeat}} \\
        \cline{2-13} & $G$ & $G'$ & $G$ & $G'$ & $G$ & $G'$ & $G$ & $G'$ & $G$ & $G'$ & $G$ & $G'$ \\
        \hhline{|=||=|=|=|=|=|=|=|=|=|=|=|=|} 
               $|V|$             & 1000 & 1000 & 1998 & 1998 & 1001 & 1001 & 1999 & 1999 & 9765 & 9765 & 12645 & 12645 \\
        \hline $|E|$             & 1000 & 1 & 26099 & 25100 & 1999 & 1000 & 27007 & 25952 & 10324 & 1111 & 67053 & 58348 \\
        \hline Density           & 0.001001 & 0.000001 & 0.006541 & 0.006290 & 0.001997 & 0.000999 & 0.006762 & 0.006498 & 0.000108 & 0.000012 & 0.000419 & 0.000365 \\
        \hline Average Degree    & 2 & 0.002 & 26.125 & 25.125 & 3.994 & 1.998 & 27.021 & 25.965 & 2.114 & 0.228 & 10.605 & 9.229 \\
        \hline Maximum Degree    & 999 & 1 & 999 & 73 & 1000 & 1000 & 1056 & 1052 & 9213 & 41 & 8705 & 2719 \\
        \hline Harmonic Diameter & 545.206 & 999000 & 9.345 & 9.374 & 500.750 & 1001 & 3.115 & 5.526 & 22.903 & 54947.822 & 4.431 & 9.329 \\
        \hline 
    \end{tabularx}
    \caption{Dataset and canary pair details}
    \label{tab:dataset}
\end{table*}

\subsection{Results and Discussion}
\begin{figure*}[t]
    \centering
    \includegraphics[width=0.48\linewidth, trim=0.25cm 0.25cm 0.25cm 0cm, clip]{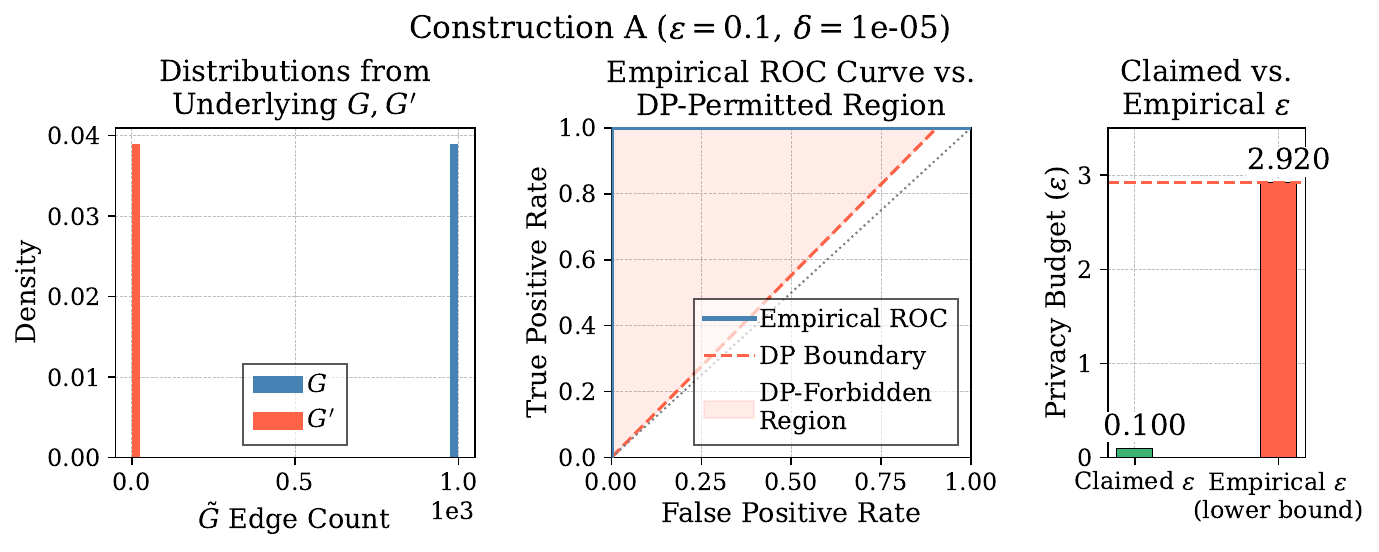} \hfill
    \includegraphics[width=0.48\linewidth, trim=0.25cm 0.25cm 0.25cm 0cm, clip]{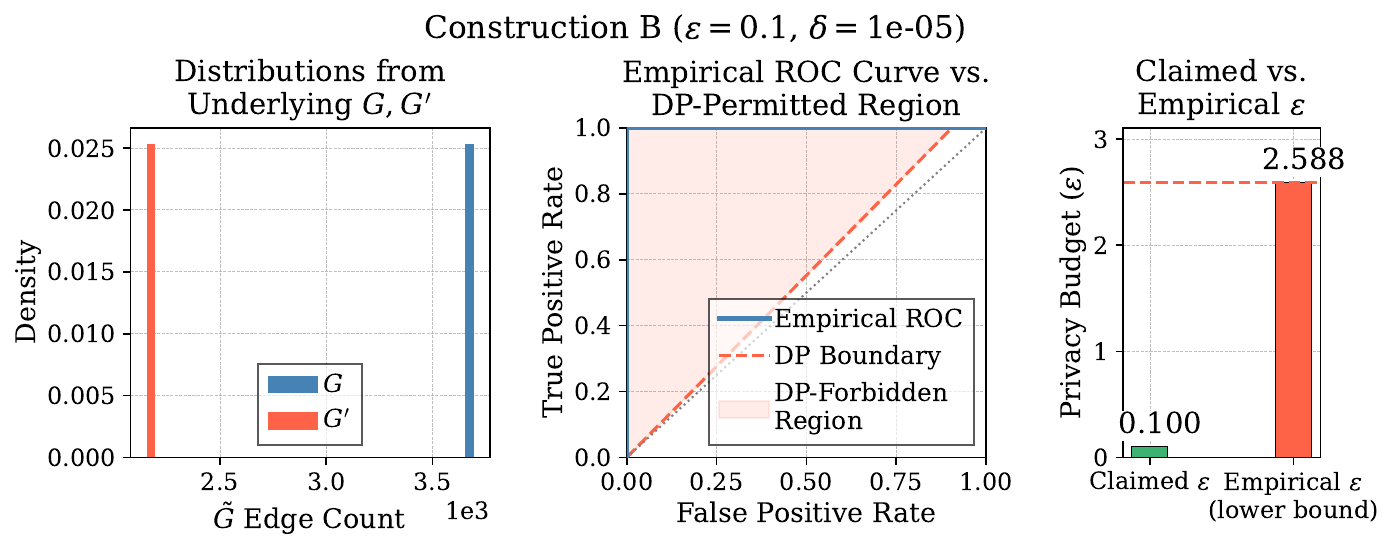} \\
    \vspace{8pt}
    \includegraphics[width=0.48\linewidth, trim=0.25cm 0.25cm 0.25cm 0cm, clip]{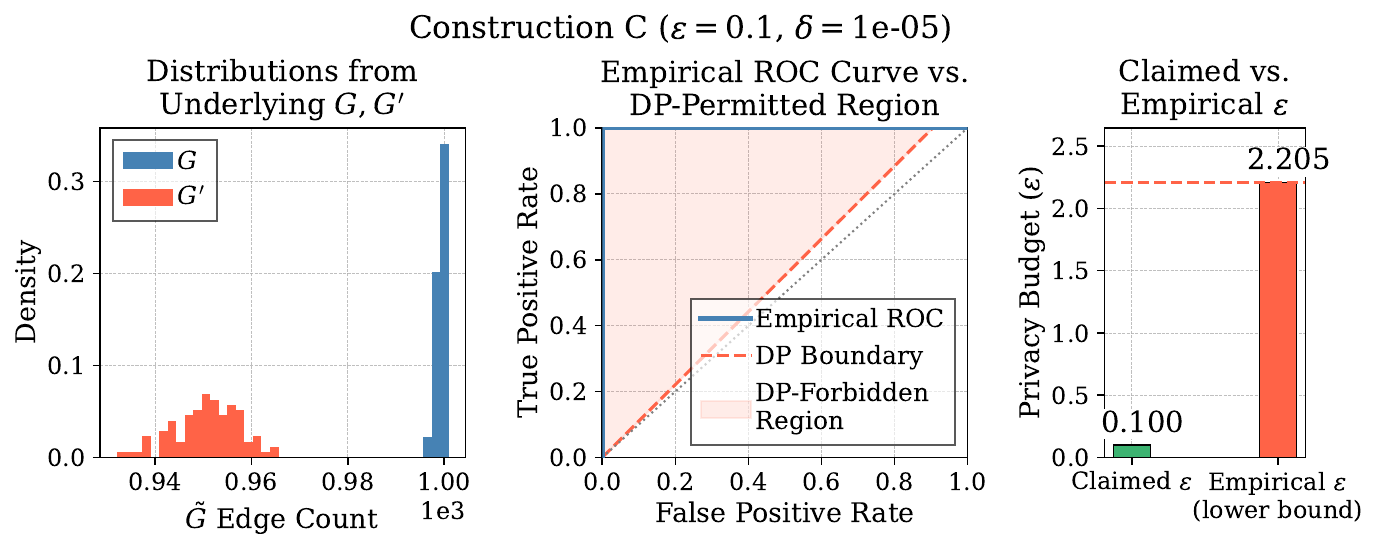} \hfill
    \includegraphics[width=0.48\linewidth, trim=0.25cm 0.25cm 0.25cm 0cm, clip]{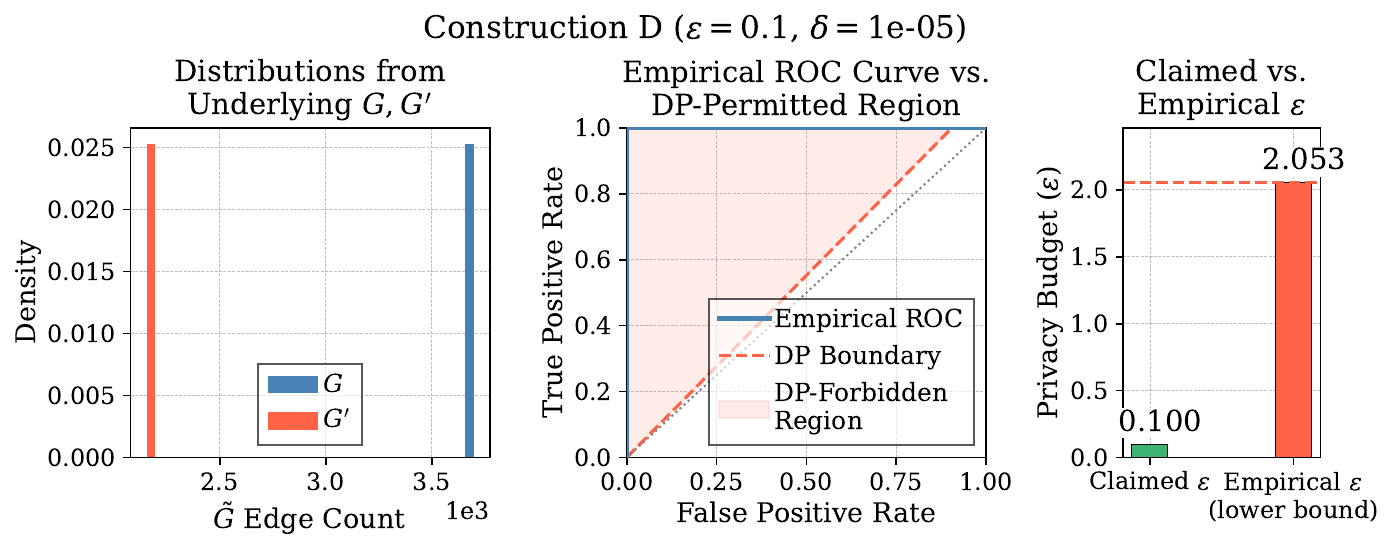} \\
    \vspace{8pt}
    \includegraphics[width=0.48\linewidth, trim=0.25cm 0.25cm 0.25cm 0cm, clip]{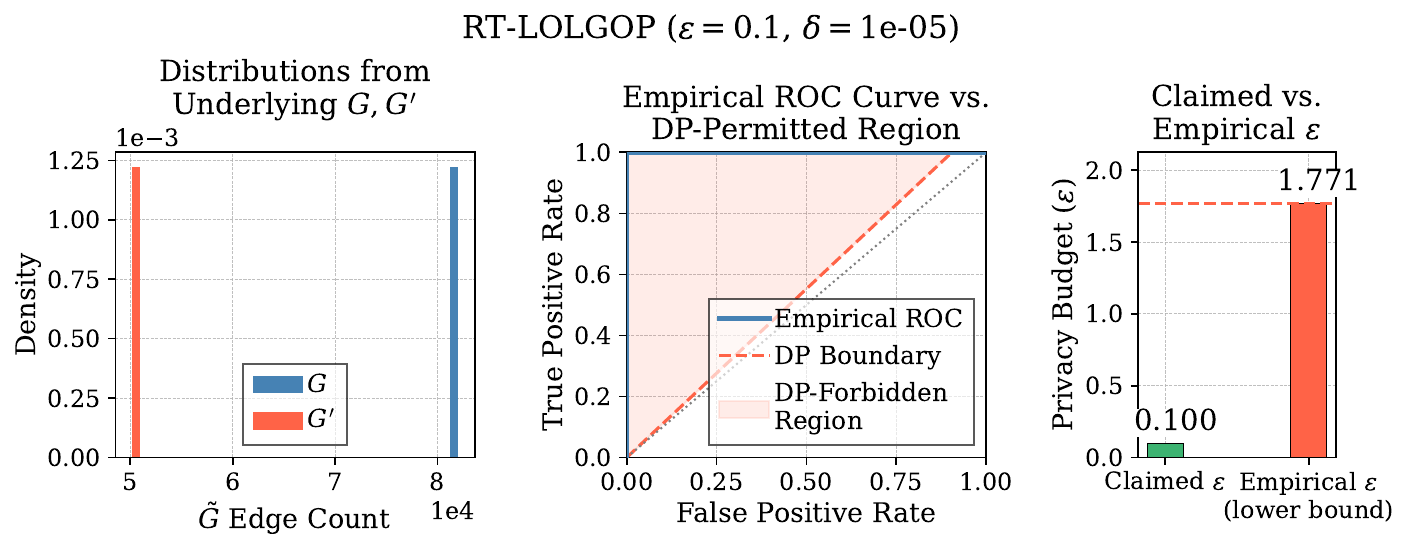} \hfill
    \includegraphics[width=0.48\linewidth, trim=0.25cm 0.25cm 0.25cm 0cm, clip]{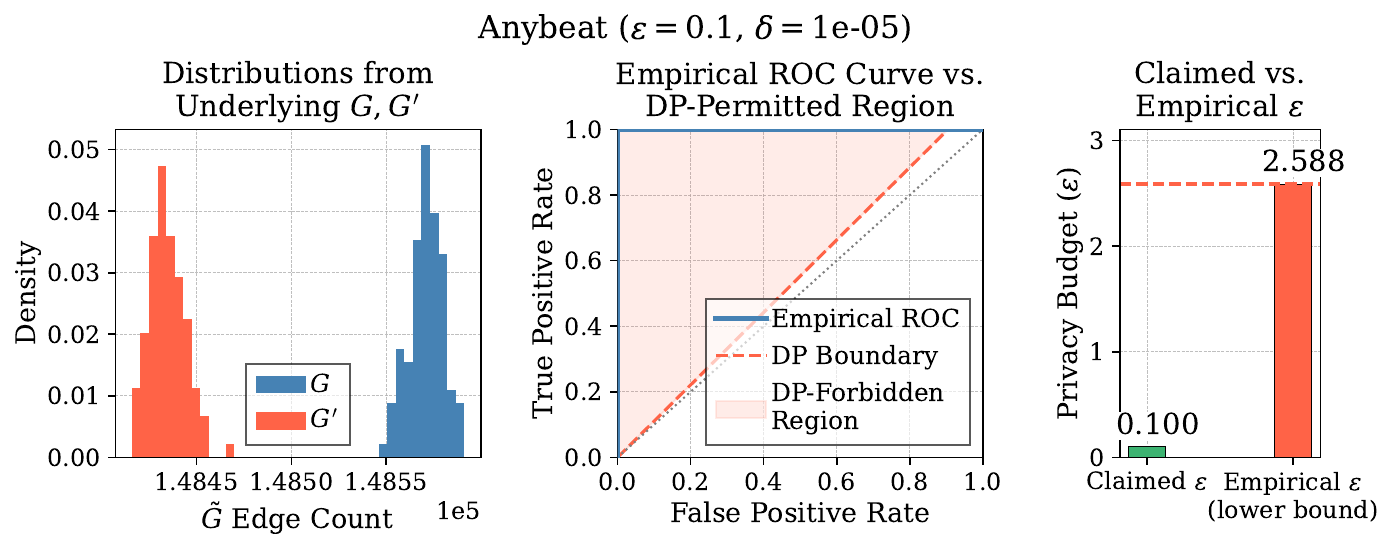} \\
    \caption{Audit results across all canary pairs using edge count as the witness statistic. Each pair yields a counterexample to the claimed DP guarantee, certified at the individual 99\% confidence level.}
    \label{fig:audit}
\end{figure*}

Findings for the edge count witness statistic on each of the six audited canary pairs are shown in Fig.~\ref{fig:audit}. Across the synthetic constructions and real-world canaries, we observe clear separations between the output distributions induced by the neighbouring inputs, indicating that the released graphs retain information about the underlying dataset. In each presented case, the certified lower bound on $\varepsilon$ exceeds mechanism's claimed value. Thus, each canary pair shown yields a statistically certified counterexample to the claimed $(\varepsilon,\delta)$ guarantee at the individual 99\% confidence level.

Importantly, the violation persists under progressively stronger restrictions on the graph domain, including non-empty, then connected graphs. This demonstrates that the observed behaviour is not confined to degenerate graph constructions. Furthermore, the progression from Constructions A and C to Constructions B and D reduces the relative influence of the perturbed node on the overall graph. In particular, Constructions B and D embed the canary within a substantially larger background structure such that the induced change represents less than 10\% of the graph's edge count. The persistence of certified violations under these conditions suggests that the observed behaviour is not solely a consequence of the perturbation dominating the graph.

The same qualitative effect is observed on canaries derived from real-world datasets, suggesting that the detected violation is not solely an artefact of controlled synthetic examples. Since DP requires the guarantee to hold for every neighbouring pair and every measurable event, a single certified counterexample is sufficient to falsify the universal privacy claim.

\begin{figure}
    \centering
    \includegraphics[width=\linewidth, trim=0.25cm 0.25cm 0.25cm 0cm, clip]{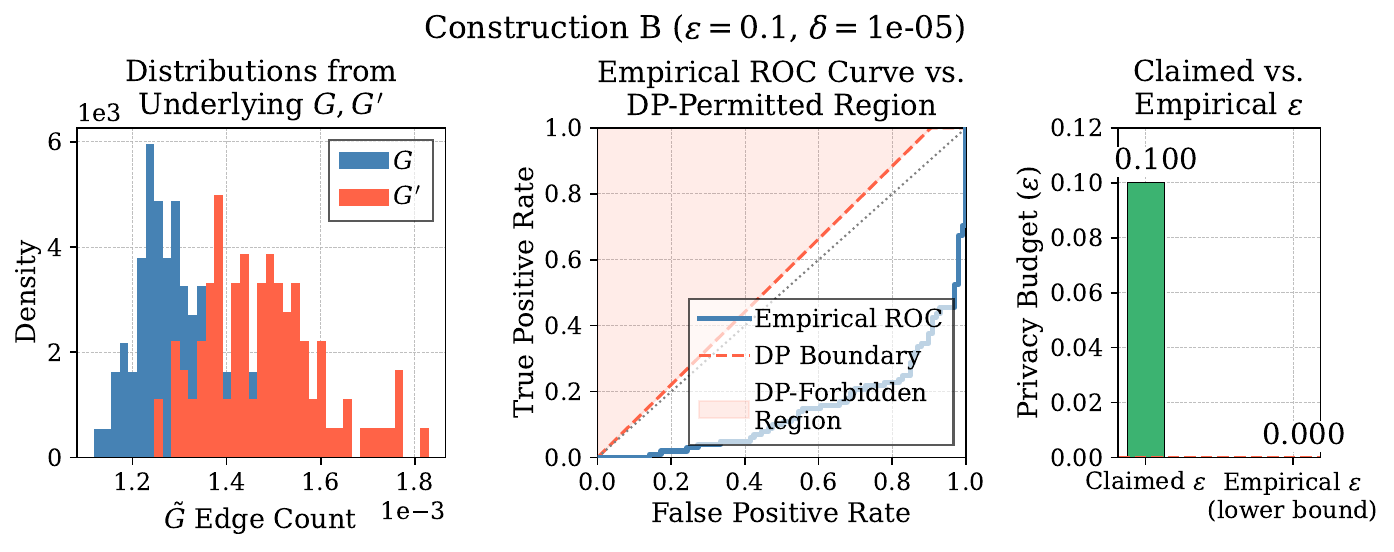}
    \caption{Audit result for Construction~B using maximum PageRank as the witness statistic. The result is inconclusive; no violation is certified.}
    \label{fig:pr-audit-unfalsified}
\end{figure}

\begin{figure}
    \centering
    \includegraphics[width=\linewidth, trim=0.25cm 0.25cm 0.25cm 0cm, clip]{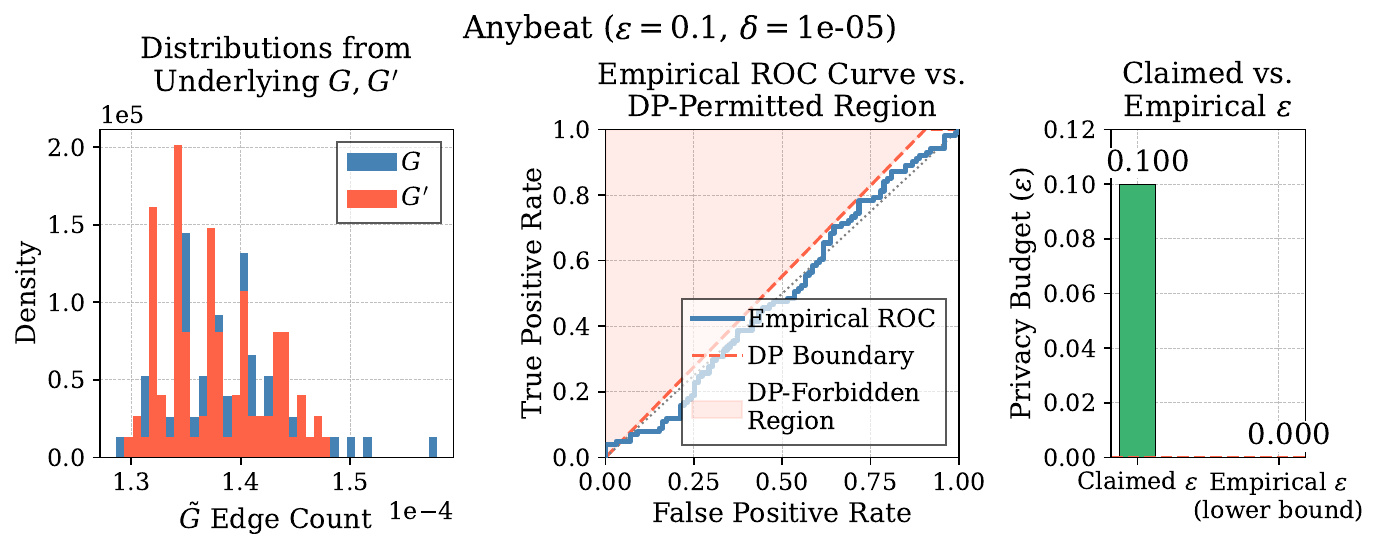}
    \caption{Audit result for Anybeat using the maximum degree as the witness statistic. The result is inconclusive; no violation is certified.}
    \label{fig:deg-audit-unfalsified}
\end{figure}

We obtained the preceding results using edge count as the witness statistic. To illustrate the importance of witness selection, Fig.~\ref{fig:pr-audit-unfalsified} reports the audit outcome for Construction~B using the maximum PageRank score~\cite{Page99} as the witness, and Fig.~\ref{fig:deg-audit-unfalsified} reports the audit outcome for Anybeat using the maximum degree as the witness. In these cases, the certified lower bound does not exceed the claimed privacy budget and the output distributions exhibit substantially weaker separation. As discussed in Sec.~\ref{audit}, such results neither falsify nor validate the claimed DP guarantee. Restricting the audit to either of these non-falsifying witnesses would therefore have missed a violation detected by edge count on the same pair. Thus, these results highlight the importance of considering multiple structural properties, rather than committing to a single witness whose effectiveness is unknown in advance.

Overall, these findings establish multiple statistically certified counterexamples to PrivDPR's bounded Node-DP claim, while also showing that violation detection depends strongly on witness selection. The results therefore support both the conclusion that the audited implementation does not satisfy its claimed privacy guarantee, and the methodological argument that graph DP audits should search across multiple witness statistics rather than relying on any single predefined metric.

\section{Limitations and Future Work}\label{future-work}
We identify the following limitations and potential avenues for future work. As with empirical DP audits more broadly, our graph-specific methodology is a falsification procedure rather than a verification procedure. An unsuccessful audit does not establish that a mechanism satisfies its claimed privacy guarantee. It only indicates that the evaluated canary, witness statistic, and decision rule did not expose a violation~\cite{Jagielski20, Nasr21, Ding18}.

In our case study, we do not perform a systematic study of audit power as a function of canary degree or other structural characteristics. Instead, our evaluation focuses on progressively restricting the graph domain and reducing the relative influence of the perturbed structure. A more comprehensive investigation of how canary topology affects audit outcomes remains an interesting direction for future work. Furthermore, our case study focuses on a single Node-DP mechanism, thus does not characterise the prevalence of similar issues across the broader graph privacy literature. 

Future work could apply the proposed methodology to additional classes of graph privacy mechanisms, including Edge-DP methods, graph analytics mechanisms, and private graph learning pipelines. It would also be valuable to investigate automated approaches to canary construction and witness statistic selection in the graph setting, as well as integrating empirical auditing into broader graph privacy evaluation frameworks.

\section{Conclusion}\label{conclusion}
We presented a methodology for empirically auditing DP graph release mechanisms. Unlike traditional auditing settings, graph auditing requires explicit consideration of both the structural context in which neighbouring perturbations are embedded and the witness statistics used to expose resulting distributional differences. To address these challenges, we proposed a methodology that separates graph canary construction from witness statistic selection while retaining statistically certified privacy testing. 

We implemented the methodology as an open-source Python toolkit and applied it to the released implementation of PrivDPR, a state-of-the-art Node-DP graph release mechanism. Through a collection of synthetic and real-world canary pairs, we obtained certified privacy lower bounds that exceed the claimed privacy budget, thereby producing counterexamples sufficient to falsify the claimed guarantee for the audited implementation. Importantly, the observed violations persisted under progressively stronger restrictions on the input domain, including non-empty and connected graphs, and were also observable on real-world datasets. 

Our experiments also demonstrated that different witness statistics can vary substantially in their ability to expose privacy violations. This finding reinforces the need to evaluate a diverse battery of statistics rather than relying on any single predefined metric in the graph setting. More broadly, the results highlight the value of empirical auditing as a complement to formal privacy analyses when evaluating DP graph release mechanisms. While auditing cannot certify privacy, it can identify implementation-level behaviour that is inconsistent with claimed guarantees and provide an additional layer of scrutiny for deployed graph privacy mechanisms.

We aim for the proposed methodology and accompanying toolkit to encourage the routine inclusion of empirical auditing within the evaluation of future graph DP mechanisms.

\begin{acks}
This research was supported by the Commonwealth through an Australian Government Research Training Program Scholarship.\footnote{DOI: \href{https://doi.org/10.82133/C42F-K220}{10.82133/C42F-K220}} The authors also thank UNSW and the anonymous reviewers.
\end{acks}

\section*{Ethics and Privacy Statement}
The experiments in this work use synthetic graphs and publicly available datasets sourced from the Network Data Repository~\cite{nr}. No participants were recruited or contacted. The case-study audit used only publicly released artefacts and documentation, with no access to non-public information.

\section*{AI Use}
AI-based tools were used to assist with proofreading, improving the clarity and flow of the text, and to assist with formatting in \LaTeX{}. AI-based tools were not used as part of the research methodology or for generating scientific claims, results, analyses, or conclusions. The authors have manually verified and take full responsibility for the accuracy, originality, and integrity of the output of all AI-based tools used in this work.

\bibliographystyle{ACM-Reference-Format}
\bibliography{references}

\appendix

\end{document}